\documentclass[12pt]{article}
\usepackage{setspace,picins}
\usepackage[]{graphicx}

\def\gta{\ifmmode {\mathbin{\lower 3pt\hbox   
    {$\,\rlap{\raise 5pt\hbox{$\char'076$}}\mathchar"7218\,$}}}
    \else {${\mathbin{\lower 3pt\hbox
    {$\rlap{\raise 5pt\hbox{$\char'076$}}\mathchar"7218\,$}}}
    $}\fi}
\def\lta{\ifmmode {\,\mathbin{\lower 3pt\hbox   
    {$\,\rlap{\raise 5pt\hbox{$\char'074$}}\mathchar"7218\,$}}}
    \else {${\mathbin{\lower 3pt\hbox
    {$\rlap{\raise 5pt\hbox{$\char'074$}}\mathchar"7218\,$}}}
    $}\fi}

\newcommand{\apj}{ApJ}
\newcommand{\apjl}{ApJL}
\newcommand{\aap}{A\&A}
\newcommand{\prd}{PRD}
\newcommand{\mnras}{MNRAS}

\title{Decadal Survey White Paper: Probing Stellar Dynamics in 
Galactic Nuclei}
\author{M. Coleman Miller$^1$, Tal Alexander$^2$, Pau Amaro-Seoane$^3$,\\ 
Aaron J. Barth$^4$, Curt Cutler$^5$, Jonathan R. Gair$^6$, Clovis Hopman$^7$,\\
David Merritt$^8$, E. Sterl Phinney$^9$, 
and Douglas O. Richstone$^{10}$}

\date{\small }

\begin{document}

\maketitle

\pagestyle{empty}
\begin{abstract}
Electromagnetic observations over the last 15 years have
yielded a growing appreciation for the importance of supermassive
black holes (SMBH) to the evolution of galaxies, and for the intricacies
of dynamical interactions in our own Galactic center.  Here we
show that future low-frequency gravitational wave observations,
alone or in combination with electromagnetic data, will open up
unique windows to these processes.  In particular, gravitational
wave detections in the $10^{-5}-10^{-1}$~Hz range will yield
SMBH masses and spins to unprecedented precision and will provide
clues to the properties of the otherwise undetectable stellar
remnants expected to populate the centers of galaxies.  Such observations
are therefore keys to understanding the interplay between SMBHs
and their environments.
\end{abstract}

\footnotetext[1]{University of Maryland, Department of Astronomy}
\footnotetext[2]{Faculty of Physics, Weizmann Institute of Science}
\footnotetext[3]{Max Planck Institut f\"ur Gravitationsphysik 
(Albert-Einstein-Institut)}
\footnotetext[4]{Department of Physics and Astronomy, University of 
California at Irvine}
\footnotetext[5]{Jet Propulsion Laboratory, California Institute of Technology}
\footnotetext[6]{Institute of Astronomy, University of Cambridge}
\footnotetext[7]{Leiden University, Leiden Observatory}
\footnotetext[8]{Department of Physics and Center for Computational Relativity and 
Gravitation, Rochester Institute of Technology}
\footnotetext[9]{Theoretical Astrophysics, California 
Institute of Technology}
\footnotetext[10]{Department of Astronomy, University of Michigan}
\newpage

\pagenumbering{arabic}
\pagestyle{plain}

\begin{spacing}{0.90}

\section{Introduction}

This white paper is directed to the Galactic Neighborhood
panel of the Decadal Survey.  Within the topics of the panel
we identify four fundamental questions for the next decade:

\begin{itemize}

\item How are dark matter and normal matter structured in the
local group?

\item How does the environment influence the evolution of the distribution
of stars and gas, in position and velocity, in galaxies?

\item How are matter, energy, and magnetic fields exchanged between
stars and interstellar gas, or between galaxies and the
intergalactic medium?

\item How do black holes affect the evolution and environments of galaxies 
and the universe, and how do galaxies and their evolution affect the 
population of black holes?

\end{itemize}

Related to the last question, in the last $\sim 10-15$ years it
has become progressively more evident that, far from being
passive receptacles of matter in galactic centers, supermassive
black holes have a key role in driving the evolution of galaxies
and galaxy clusters.  It is therefore important to conduct a
variety of observations and theoretical simulations of galactic
nuclear dynamics to evaluate the interactions between black holes
and their environments.  Here we demonstrate that detections of
low-frequency gravitational radiation from inspirals of
stellar-mass objects into supermassive black holes will open a
unique window into stellar dynamics in galactic nuclei.  Analysis
of these inspirals will allow us to measure black hole masses and
spins with unprecedented precision, and provide clues to the
properties and interactions of the otherwise invisible stellar
remnants expected in galactic nuclei.  It will also complement
work on nuclear dynamics, hyper-velocity stars, and other
electromagnetic observations. Note: this white paper is also
being submitted to the ``Galaxies Across Cosmic Time" panel, due
to the information that could be obtained about galaxies out to
redshifts $z\sim 1$.

\section{Properties and evolution of supermassive black holes}

Supermassive black holes (SMBHs) accreting gas in galactic nuclei
were first proposed in the 1960s \cite{Sal64,ZN64} to explain the
enormous luminosities of the newly-discovered quasars. Dynamical
measurements in the last decade and a half have verified
Lynden-Bell's (1969) suggestion that SMBHs are present even in
many quiescent, giant galaxies,  and indeed for black hole masses
$M_{\rm BH}\gta 10^7~M_\odot$ (bulge luminosity $\gta
10^{10}L_\odot$) the SMBH mass is correlated with the velocity
dispersion $\sigma$ of the bulge  via the $M-\sigma$ relation:
$M_{\rm BH}\propto \sigma^{4-5}$ (e.g., \cite{FF05}). This
suggests that, at least for large SMBHs, black hole growth and
galaxy evolution are tightly coupled.

Black hole mass measurements are, however, challenging, especially
for lower-mass SMBHs.  The primary source of black hole masses
nearby is direct observations and modeling of motions of stars and
gas orbiting the centers of galaxies coupled with dynamical
models.  These measurements are always limited by the spatial
resolution of the observations, by degeneracy between the
orbital distribution functions and BH mass, and by degeneracy
between the mass of stars in the galactic center and the BH mass.
These limitations are not likely to be greatly reduced in the next
decade. With a few exceptions, the better measurements remain
uncertain by a factor of two, and some are only good to a factor of
ten.  Methods such as reverberation mapping or those based on
correlations between emission line width and mass are calibrated
by the direct dynamical measurements and are thus even more
uncertain.  Indeed, even the number
density of the galaxies that would host black holes in the $\sim
10^{5-7}~M_\odot$ range is uncertain by at least an order of
magnitude, and depends on the estimator used (e.g., \cite{AR02,GH07}).

In a parallel fashion, we understand the growth of large SMBHs because
comparisons of the current mass density of SMBHs in the universe
(dominated by masses around $10^8~M_\odot$) with the integrated light
from quasars (also dominated by central engines with $M_{\rm BH}\sim
10^8~M_\odot$) imply that most of their mass has been acquired via
radiatively efficient gas accretion \cite{Soltan82,Marconi04}.  In
contrast, we know comparatively little about how SMBHs grow to $M_{\rm
BH}\sim 10^{5-7}~M_\odot$ (a range that obviously includes the SMBH at
the center of our own Milky Way, at $\sim 4 \times 10^6 M_\odot$). 
This could occur via gas accretion, as for more massive black holes. 
It could also occur by accretion of multiple stars coming in from
random directions \cite{Hills75}; by mergers with stellar-mass black
holes, neutron stars, or intermediate-mass black holes
(IMBHs)\footnote{Intermediate mass black holes are hypothetical black
holes with masses in the range $10^2 M_{\odot}<M_{\bullet}<10^4
M_{\odot}$.  Currently the only evidence for their existence is
indirect, in contrast to the direct dynamical evidence for
stellar-mass and supermassive BHs.} formed in star clusters
\cite{QS90,Por06} (accretion of compact objects would not generate significant
luminosity, and would thus evade the Eddington limit and possibly
allow rapid early growth); or even by exotic mechanisms such as the
accretion of dark matter \cite{Ostriker00} or direct production from
collapse of supermassive stars \cite{ST79}.  Thus the early growth of
all SMBHs, and the history of current lower-mass SMBHs, is an open
question that has bearing on early structure formation and galaxy
evolution.

Precision measurements of the masses of low-mass SMBHs are
required to untangle their evolutionary processes. During the past
few years, 8m-class telescopes with adaptive optics have been
employed for a small but growing number of dynamical detections of
SMBHs in very massive, bulge-dominated galaxies
\cite{Houghton06,Nowak08}.  However, current observational
capabilities are not well suited to detection of lower-mass black
holes in the nuclei of disk-dominated or dwarf galaxies, due to
the smaller gravitational radius of influence of low-mass black
holes, and as indicated above it is likely that future mass
uncertainties using these methods will still be at least a factor
of two even with the next generation of large ($\sim25-30$m)
ground-based telescopes.  As an alternate method, as we discuss
below detection of low-frequency gravitational waves from extreme
mass ratio inspiral events (EMRIs) will determine the mass of the
primary BH to a relative precision of about $10^{-4}$, better than
the best current measurements. Although small in number, these
measurements have great potential: even though it will probably
be rare that an EMRI is observed from near an SMBH whose mass has
been estimated by electromagnetic methods, in a statistical
sense the mass distribution obtained using EMRIs could expose unrecognized
errors or improve the confidence in the current estimates, and
they might point the way toward improvements in the  current suite
of techniques.

Spin measurements are also important because they allow us to
distinguish between prograde gas accretion (which is expected to
spin the SMBHs up to values of $a/M=cJ/GM^2\gta 0.9$;
\cite{Bardeen70}), mergers with comparable-mass black holes (which
characteristically yield $a/M\sim 0.7$ in many cases;
\cite{Baker04}), and accretion of many masses with uncorrelated
direction (e.g., stars or compact objects), which is expected to
produce $a/M\ll 1$ \cite{Young77}. It is currently possible in some
cases to measure spin using X-ray observations of  Fe~K$\alpha$
profiles \cite{RF08}, but these are uncommon and the typical
uncertainties are significant: often $|\Delta(a/M)|\gta 0.1$ unless
the spins are near maximal \cite{Bre07}.  There is also an active
current discussion about the assumptions that underlie this method
of spin determination (e.g., if emission can extend somewhat inside
the innermost stable circular orbit, this would be interpreted as a
spin greater than the actual one).

\section{The Promise of Low-Frequency Gravitational Waves}

In contrast to the electromagnetic methods described above,
low-frequency space-based gravitational-wave detectors, which will focus
on the band $10^{-4}~{\rm Hz} < f < 10^{-1}~{\rm Hz}$, 
are particularly sensitive to
gravitational waves from SMBHs in the range ~$10^5-10^7 M_{\odot}$. 
Whereas main-sequence stars that are captured by such SMBHs are tidally
disrupted before they pass through the SMBH's event horizon,
stellar-mass compact objects (black holes, neutron stars, and white
dwarfs) get swallowed whole. A reasonable fraction of such compact
objects will slowly spiral into the SMBH (as opposed to be being
swallowed directly, on the first ``pass" by the SMBH).  The slowly
inspiraling compact objects will have gravitational wave frequencies of
$f>10^{-4}$~Hz years before they are ultimately swallowed, and will
hence undergo $\sim 10^5$ orbits at these frequencies. These inspirals
will be detectable to cosmological distances, with detection rates
estimated at tens per year \cite{Gai08}. Of the three classes of
compact object, BH inspirals are expected to dominate the detection
rate, both because higher mass means greater GW amplitude (and hence a
larger detection volume, out to $z\sim 1$) and because mass
segregation  concentrates the heaviest objects closest to the SMBH.

The highly relativistic orbits of EMRIs, lying within $\sim 5-10$
Schwarzschild radii of the SMBH, display extreme versions of both
relativistic pericenter precession and Lense-Thirring precession
of the orbital plane about the SMBH's spin axis (see Figure~1).
The large number of cycles and complexity of the orbits
encode wonderfully detailed information concerning the system's
physical parameters.  The mass of the compact object, the mass and spin of
the SMBH, and the eccentricity of the orbit (at some fiducial
instant--say 6 months before plunge), will typically all be
determined to  fractional accuracy $\sim 10^{-4}$.  The orbit's
inclination with respect to SMBH spin will be measured to $\Delta
(\rm cos)\iota \sim 10^{-3} -10^{-2}$ \cite{Bar04a,HG08}.  Analysis
of these orbits will also yield precise tests of the predictions
of general relativity in strong gravity, in a way that is complementary
to electromagnetic observations with proposed projects such as 
GRAVITY \cite{Eisen08}.
Any IMBHs captured by SMBHs would be observable out to very
high redshift; e.g., $10^{2-3} M_{\odot}$ IMBHs spiraling into SMBHs with
$M(1+z)$ in the range $3\times 10^5-3\times 10^6 M_{\odot}$ could be detected
out to $z=20$, with the IMBH mass typically determined to better than
$1\%$ (and significantly better than that for redshifts $z$ less than a
few).  Such observations could produce the first definitive
evidence of the existence of IMBHs.
These precise measurements will be invaluable for studies of the
many dynamical processes expected in galactic nuclei, which we
now discuss.

\parpic[r][r]{\mbox{\resizebox{0.45\hsize}{!}
        {\includegraphics{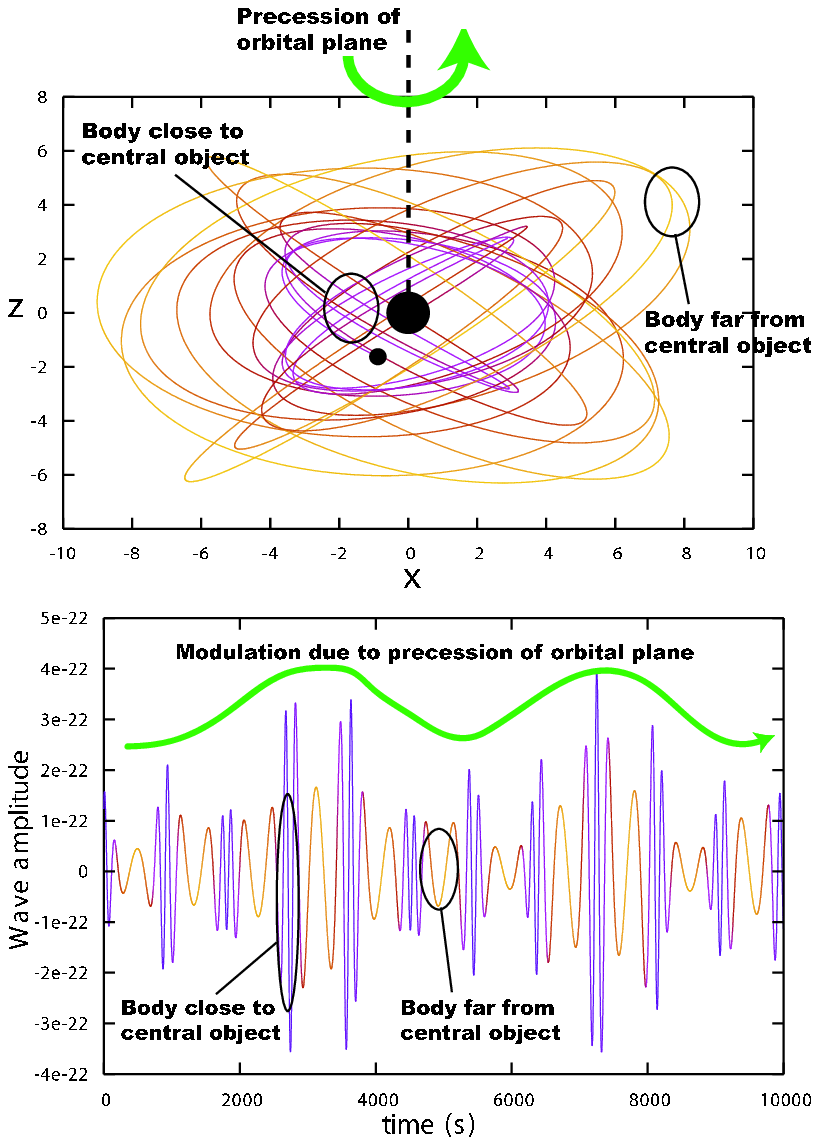}}}}

\vspace{0.3cm}
\noindent
Cartoon of an EMRI orbit, as viewed from the side (top panel) and emitted
gravitational wave (bottom panel). The gravitational wave is characterized
by higher amplitude and frequency radiation associated with extreme
pericenter precession when the body is close to the central object, and
lower amplitude and frequency radiation when the body is further away. There
is an overall modulation due to precession of the orbital plane. The
waveform is colored to illustrate this structure. (Taken from J Gair, Phil.
Trans. Roy. Soc. A366, 4365 (2008)).

\vspace{1.5cm}

\section{Galactic center dynamics and extreme mass ratio inspirals}

\subsection{Observations of the Galactic center}

Extensive near-infrared and X-ray observations of the inner parsec of
the Galactic center reveal a remarkably detailed and surprising
picture.  Components include a $4\times10^6 M_{\odot}$ black hole
\cite{Gil08} embedded in an extended population of old, relaxed stars
\cite{Sch07}, an isotropic cluster of seemingly normal young hot stars
within $\rm{few}\times0.01$ pc from the SMBH, very massive young stars
orbiting coherently in a disk \cite{Lev03, Bar08} (possibly formed
from a fragmenting accretion disk), and a few X-ray point sources
\cite{Mun05}.

The origin, evolution, and physical processes governing this system
remain mysterious; e.g., the
spin of the SMBH is completely unknown, and the population of dark
compact objects around the SMBH is also unknown.
Theory predicts that  the central parsec harbors
$O(10^4)$ stellar black holes (BHs) that sank there over the
Galaxy's lifetime \cite{Mor93, Fre06, Hop06b, Ale08},
together with the stellar BHs that are believed to be produced
locally by the unusual mode of massive star formation in a disk.
This hypothesized cluster of black holes dominates the dynamics of
the inner $\sim0.01$~pc of the galaxy, and interacts with gas and
stars there. For example, it likely drives rapid resonant relaxation
\cite{Rau98}. The existence of such a cluster cannot yet be
dynamically confirmed \cite{Mou05, Ghe08}; in general, very little
is  known empirically about the birth and mass functions of stellar
BHs.  Given their important role for the dynamics of regions close
to SMBHs, this  represents a significant gap in our understanding of
galactic nuclei.  Conversely, the Galactic center, which harbors up
to $10^{-3}$ of all Galactic stellar BHs in only $\sim10^{-10}$ of
the Galactic  volume, provides a unique opportunity to study the
properties of stellar BHs.  In addition, the Galactic center may
contain several intermediate-mass black
holes \cite{Han03, Por06, Mer08}.

\subsection{How EMRI observations of other galaxies will probe 
galactic dynamics}

The puzzles posed by the center of our Galaxy and others couple 
long-standing key
questions in stellar dynamics, gas dynamics, star formation and
stellar evolution. At the same time, these systems offer exciting
prospects for significant progress because of the wealth of data
available on complex structures strongly constrained by the extreme
environment. In particular, the presence of so many stellar BHs in the
vicinities of SMBHs in galactic centers makes it possible to uniquely
combine the powers of high-precision electromagnetic observations with
those of low-frequency gravitational radiation from EMRIs, which will
place constraints on the stellar contents and dynamics.

Specific examples include:
\begin{itemize}

\item Mass segregation will drive many stellar black holes towards the
center. As a result, the rate at which EMRIs will be detected via
gravitational waves is estimated to be roughly $10^{-7}$ per galaxy
per year.  For an instrument such as the Laser Interferometer
Space Antenna (LISA), which will probe out
to a redshift $z\sim 1$, the net detection rate is expected to
be tens to hundreds per year \cite{Gai08}. 
The eccentricities and inclinations of EMRIs in the $f>10^{-4}$~Hz
gravitational wave band will be signatures of their origin through
processes such as two-body scattering (for a recent review see
\cite{Gai08}), tidal separation of binaries (\cite{Mil05}; see
Figure~2), or settling
of stellar-mass black holes via repeated interaction with an accretion
disk  \cite{Mir05,Lev07}.  The effects of tidal separation may already
have been seen, as this process is the leading candidate to explain
the so-called hypervelocity stars observed escaping from our Galaxy
(e.g., \cite{Brown09}).  Combining gravitational wave and electromagnetic
observations is key to understanding and interpreting
stellar populations there.

\item Discovery of EMRIs will
provide unique information about the mass spectrum of
stellar black holes in galactic nuclei, in particular their upper
mass limit. This is key for understanding the formation of stellar BHs
and their relation to their progenitors.

\item Detection of EMRIs will also give the
distribution of the SMBH spins for SMBHs of masses up to ${\rm few}
\times 10^6 M_{\odot}$ \cite{Bar04a}, and thus help to disentangle the
formation history of SMBHs.

\item The detection of the inspiral of an IMBH
into a SMBH will give direct evidence for the existence of IMBHs
\cite{Mil05b}, and identify a major dynamical component in 
galactic centers.

\item EMRIs involving low-mass white dwarfs spiraling into
SMBHs with $M\lta 10^5~M_\odot$ may yield a strong
and extended electromagnetic outburst due to the tidal destruction
of the white dwarf and subsequent accretion of gas \cite{Ses08}.
For more on this process, please read the white paper on tidal
disruptions led by Suvi Gezari.

\end{itemize}

\parpic[r][r]{\mbox{\resizebox{0.45\hsize}{!}
        {\includegraphics{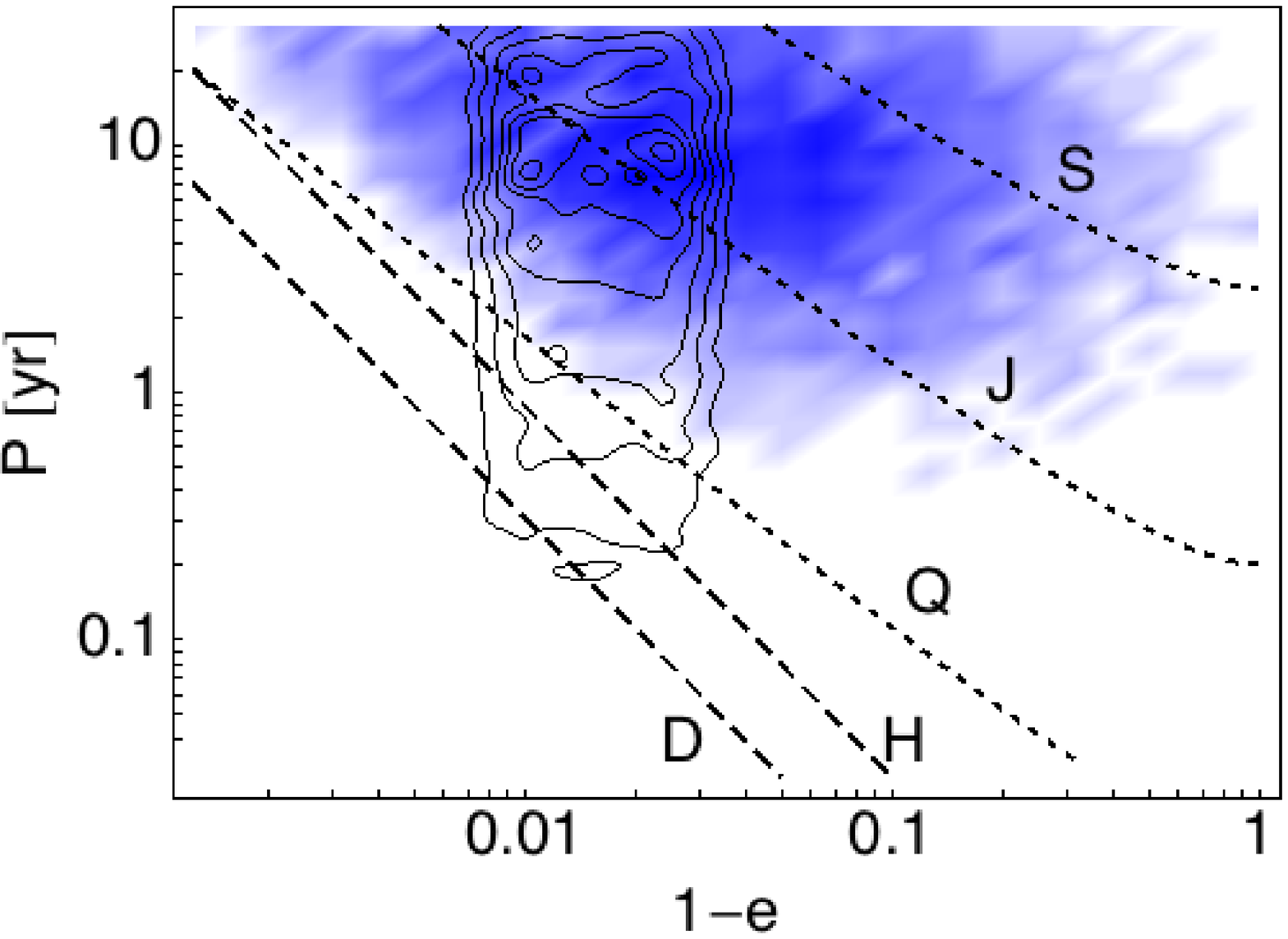}}}}

\noindent
Simulations of binary disruption capture and subsequent orbital
evolution of stars on highly relativistic orbits around the Galactic
MBH (T. Alexander 2009, private communication). 

\noindent
The initial period and eccentricity (contour lines) evolve over 10 Myr due to
relaxation, GW emission, tidal interactions and collisions with compact
remnants. Of the 40\% of stars that survive destruction (blue area) by tidal
heating (H) or disruption (D) (dashed lines), many are found on tight eccentric
orbits (below the plotted lines), where relativistic precession ($>5\mu$as/yr)
due to the Schwarzschild periapse shift (S), frame dragging (J), and the MBH
quadrupole moment (Q) can be detected by high-precision IR interferometry and
used to measure the MBH spin and test GR \cite{Will08}.

\section{Summary}

Observations of extreme mass ratio inspirals via low-frequency
gravitational radiation will yield unprecedented precision in
the measurements of the masses and spins of supermassive black
holes, and of the masses of the stellar-mass objects that spiral
into them.  They may provide definitive evidence of the existence
of intermediate-mass black holes, and the properties of EMRIs
will allow us unique glimpses into stellar dynamics near 
SMBHs.  As a result, especially when combined with future
electromagnetic observations of galactic nuclei, low-frequency
gravitational wave observations will play a key role in determining 
how black holes affect the evolution and environments of galaxies 
and the universe, and how galaxies and their evolution affect the 
population of black holes.

\end{spacing}


\begin{thebibliography}{99}
\expandafter\ifx\csname natexlab\endcsname\relax\def\natexlab#1{#1}\fi
\parskip=-0.05in

\bibitem{Ale08}
{Alexander}, T., \& {Hopman}, C. 2008, ArXiv e-prints (arXiv:0808.3150)

\bibitem{AR02}
{Aller}, M.~C., \& {Richstone}, D. 2002, ApJ, 124, 3035

\bibitem{Baker04} Baker, J., Campanelli, M., Lousto, C.~O., \&
Takahashi, R.\ 2004, \prd, 69, 027505

\bibitem{Bar04a}
{Barack}, L., \& {Cutler}, C. 2004, \prd, 69, 082005

\bibitem{Bardeen70} 
Bardeen, J.~M.\ 1970, Nature, 226, 64 

\bibitem{Bar08} {Bartko}, H., et al.\ 2008, ArXiv e-prints (arXiv:0811.3903)

\bibitem{Bre07} {Brenneman}, L.~W. 2007, PhD thesis

\bibitem{Brown09} Brown, W.~R. et al. 2009, ApJ, 690, 1639

\bibitem{Eisen08} {Eisenhauer}, F. et al. 2008, 2008 SPIE proceedings,
vol. 7013, p. 70132A

\bibitem{FF05}
{Ferrarese}, L., \& {Ford}, H. 2005, Spa. Sci. Rev., 116, 523-624

\bibitem{Fre06}
{Freitag}, M., {Amaro-Seoane}, P., \& {Kalogera}, V. 2006, \apj, 649, 91

\bibitem{Gai08}
{Gair}, J.~R. 2008, ArXiv e-prints (arXiv:0811.0188)

\bibitem{Ghe08} {Ghez}, A.~M., et al.\ 2008, in IAU Symposium,
Vol. 248, 52--58

\bibitem{Gil08}
{Gillessen}, S., et al.\ 2008, ArXiv e-prints (arXiv:0810.4674)


\bibitem{GH07}
{Greene}, J.~E., \& {Ho}, L.~C. 2007, ApJ, 667, 131

\bibitem{Han03}
{Hansen}, B.~M.~S., \& {Milosavljevi{\' c}}, M. 2003, \apjl, 593, L77

\bibitem{Hills75} 
Hills, J.~G.\ 1975, Nature, 254, 295 

\bibitem{Hop05}
{Hopman}, C., \& {Alexander}, T. 2005, \apj, 629, 362

\bibitem{Hop06b}
---. 2006, \apjl, 645, L133

\bibitem{Houghton06} Houghton, R.~C.~W., et al.\ 2006, \mnras, 367, 2 

\bibitem{HG08} Huerta, E.~A., \& Gair, J.~R. 2008, arXiv:0812.4208

\bibitem{Lev07}
{Levin}, Y. 2007, \mnras, 374, 515

\bibitem{Lev03}
{Levin}, Y., \& {Beloborodov}, A.~M. 2003, \apjl, 590, L33

\bibitem{LB69} 
Lynden-Bell, D.\ 1969, Nature, 223, 690

\bibitem{Marconi04} 
Marconi, A., et al.\ 2004, \mnras, 351, 169 

\bibitem{Mer08} {Merritt}, D., {Gualandris}, A., \& {Mikkola},
S. 2008, ArXiv e-prints (arXiv:0812.4517)

\bibitem{Mil05b}
{Miller}, M.~C. 2005, \apj, 618, 426

\bibitem{Mil05}
{Miller}, M.~C., {Freitag}, M., {Hamilton}, D.~P., \& {Lauburg}, V.~M. 2005,
  \apjl, 631, L117

\bibitem{Mir05}
{Miralda-Escud\'e}, J., \& {Kollmeier}, J.~A. 2005, ApJ, 619, 30

\bibitem{Mor93}
{Morris}, M. 1993, \apj, 408, 496

\bibitem{Mou05}
{Mouawad}, N., et al.\ 2005, Astronomische Nachrichten, 326, 83

\bibitem{Mun05}
{Muno}, M.~P., et al.\ \apjl, 622, L113

\bibitem{Nowak08} Nowak, N., et al.\ 2008, \mnras, 391, 1629

\bibitem{Ostriker00} 
Ostriker, J.~P.\ 2000, Physical Review Letters, 84, 5258 

\bibitem{Por06}
{Portegies Zwart}, S.~F., et al.\ 2006, \apj, 641, 319

\bibitem{QS90} 
Quinlan, G.~D., \& Shapiro, S.~L.\ 1990, \apj, 356, 483 

\bibitem{Rau98}
{Rauch}, K.~P., \& {Ingalls}, B. 1998, \mnras, 299, 1231

\bibitem{RF08} 
Reynolds, C.~S., \& Fabian, A.~C.\ 2008, \apj, 675, 1048

\bibitem{Sal64}
{Salpeter}, E.~E. 1964, ApJ, 140, 796

\bibitem{Sch07} {Sch{\"o}del}, R., et al.\ 2007, \aap, 469, 125

\bibitem{Ses08} {Sesana}, A., et al. 2008, \mnras, 391,718

\bibitem{ST79} 
Shapiro, S.~L., \& Teukolsky, S.~A.\ 1979, \apjl, 234, L177 

\bibitem{Soltan82} 
So{\l}tan, A.\ 1982, \mnras, 200, 115 

\bibitem{Will08}
Will, C.~M. 2008, ApJ, 674, L25

\bibitem{Young77} 
Young, P.~J.\ 1977, \apj, 212, 227 

\bibitem{ZN64}
{Zeldovich}, Ya., \& Novikov, I. 1964, Dokl. Akad. Nauk. SSSR 158, 811

\end{thebibliography}
\end{document}